\documentclass[useAMS,usenatbib]{mn2e}

\usepackage{graphicx}
\usepackage{times}
\usepackage{natbib}

\begin{document}
 
\title[Why are central radio relics so rare?]{Why are central radio relics so rare?}
\author[F. Vazza, M. Br\"{u}ggen, R. van Weeren, A. Bonafede, K. Dolag, G. Brunetti]{F. Vazza$^{1,2}$%
\thanks{%
 E-mail: f.vazza@jacobs-university.de}, M. Br\"{u}ggen$^{1}$, R. van Weeren$^{3}$, A. Bonafede$^{1}$, K. Dolag$^{4}$, G. Brunetti$^{2}$\\
 %EndAName
$^{1}$ Jacobs University Bremen, Campus Ring 1, 28759, Bremen, Germany \\
$^{2}$ INAF/Istituto di Radioastronomia, via Gobetti 101, I-40129
Bologna, Italy\\
$^{3}$ Leiden Observatory, Leiden University, PO Box 9513, 2300 RA Leiden, The Netherlands\\
$^{4}$ University Observatory Munich, Scheinerstr. 1, D-81679 Munich, Germany}

\date{Accepted ???. Received ???; in original form ???}
\maketitle

\begin{abstract}

In this paper we address the question why cluster radio relics that are connected to shock acceleration, so-called radio gischt, have preferentially been found in the outskirts of galaxy clusters. 
By identifying merger shock waves in cosmological grid simulations, we explore several prescriptions for relating the energy dissipated in shocks to the energy emitted in the radio band. None of the investigated models produce detectable radio relics within 100-200 kpc from the cluster centre. All models cause $> 50$ per cent of the detectable relic emission at projected distances $> 800$ kpc. Central radio relics caused by shocks that propagate along the line-of-sight are
rare events for simple geometrical reasons, and they have a low surface brightness making them elusive for current instruments. Our simulations show that the radial distribution of observed relics
can be explained by the radial trend of dissipated kinetic energy in shocks, that increases with distance from the cluster centre up until half of the virial radius.

\end{abstract}

\label{firstpage}

\section{Introduction}
\label{sec:intro}

Radio relics can be divided into two main groups \citep{2004rcfg.proc..335K}: \emph{Radio gischt} are large elongated, often Mpc-sized, radio sources located in the periphery of merging galaxy clusters. In a few cases, their spatial co-location with shocks in the thermal gas has been determined (e.g. in the case of Abell 3667, \citealt{fi10}). They most likely trace shock fronts in which particles are accelerated via diffusive shock acceleration \citep[e.g.][]{be87}. Among them are double-relics with the two relics located on different sides of a cluster centre \citep[e.g., ][]{bonafede09, vanweeren09, 2007A&A...463..937V, 2006Sci...314..791B, 1997MNRAS.290..577R, bdr11, bag11}. \emph{Radio phoenices} are related to radio-loud AGN. Fossil radio plasma from a previous episode of AGN activity is thought to be compressed by a merger shock wave which boosts, both, the magnetic field inside the plasma as well as the momenta of the relativistic particles. Here we are only concerned with radio relics of the gischt type.

The sizes of relics and their separations from the cluster centre vary significantly.  Examples for radio
relics with sizes of 1~Mpc or even larger have been observed in Coma \citep{giovannini91},
Abell 2255 \citep{feretti97} and Abell 2256 \citep[see][for other
examples and references]{vw11}.

%as do Abell 2256, Abell 521, Abell 746, Abell 754, Abell 1300, Abell 2255 and Abell 2744). The cluster Abell 3667 contains two very luminous, almost symmetric
%relics with a separation of more than 3~Mpc \citep{1997MNRAS.290..577R}, as do ZwCl 2341.1+0000 \citep{vanweeren09, bonafede10}, ZwCl 0008.8+5215 (van Weeren et al. 2010), Abell 2345, Abell 1240 \citep{bonafede09}, 0217+70 (\citealt{bdr11}), CIZA J2242.8+5301 \citep{vanweeren10} and PLCK G287.0+32.9 (Bagchi et al. 2011).

Even though the sample of known radio relics is still small and incomplete (34, see Table 1), one can
start to find correlations between size, location and spectral index
of these unique sources, which can be compared to
simulations of relic formation. \cite{vanweeren09} find that on average the smaller
relics have steeper spectra. Such a trend is in line with predictions from
shock statistics derived from cosmological simulations
\citep{va10kp,skillman08, pf07,hoeft08}. They find that larger shock
waves occur mainly in lower-density regions and have larger Mach
numbers, and consequently flatter spectra. Conversely, smaller shock
waves are more likely to be found in cluster centres and have lower
Mach numbers and steeper spectra.
In this paper we explore why radio gischt relics are never observed close to the centre of galaxy clusters, using a set of high-resolution cosmological re-simulations of massive galaxy clusters at three cosmic epochs (z=0.0, z=0.3 and z=0.6).

\begin{figure*}
\begin{center}
\includegraphics[width=0.95\textwidth]{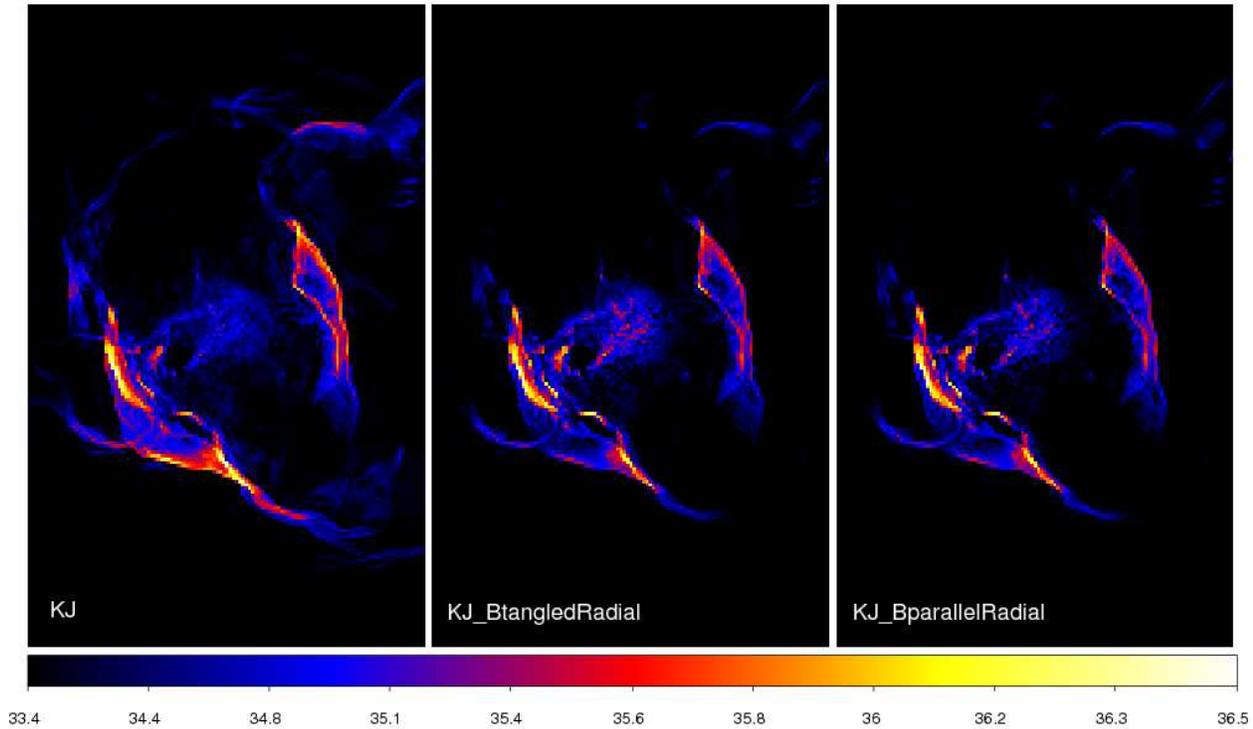}
\caption{Emitted ``radio'' power, $P$, from simulated relics (colours, in units of $\log_{\rm 10} P/$ [erg/s]), for the same galaxy cluster at z=0 and for 3 emission models (KJ,  {\it KJ\_BparallelRadial} and {\it KJ\_BtangledRadial}, in the last two models we assumed $B_{\rm 0}=5\, \mu G$). The side of each image is 10 Mpc/h.}
\label{fig:fig1}
\end{center}
\end{figure*}

\begin{table}
\caption{List of gischt radio relics. From left to right: name of host cluster, redshift and projected distance from the centre. In case of multiple relics per cluster, the letters N,S,E,W denote the direction with respect to the 
cluster centre. We notice that in the case of a few objects (labelled with "*" in the table) the identification as "radio gischt" can be debated.}
\begin{center}
\begin{tabular}{|l|l|l|}
name & z & r[kpc]  \\ \hline
Coma cluster&    0.023&       2008  \\
Abell 548b&    0.042&       693  \\
Abell 548b*&    0.042&       456  \\
Abell 3376 E&    0.0456&       845  \\
Abell 3376 W&    0.0456&       1008  \\
Abell 3667 E&    0.055&       1459  \\
Abell 3667 W&    0.055&       1887  \\
Abell 2256&    0.059&       470 \\
CIZA J0649.3+1801&    0.064&       802  \\
RXC J1053.7+5452&    0.07&       981  \\
Abell 2061&    0.078&       1558  \\
Abell 2255&    0.0806&       1065  \\
Abell 3365 E&    0.093&       1079  \\
Abell 3365 W&    0.093&       777  \\
ZwCl 0008.8+5215 E&     0.104&       798  \\
ZwCl 0008.8+5215 W&     0.104&       808  \\
Abell 1664*&     0.128&       895  \\
Abell 1240 S&     0.159&       982  \\
Abell 1240 N&     0.159&       924  \\
Abell 2345 E&     0.177&       1431  \\
Abell 2345 W&     0.177&       857  \\
Abell 1612&     0.179&       894  \\
CIZA J2242.8+5301 N&     0.192&       1565  \\
CIZA J2242.8+5301 S&     0.192&       1062  \\
Abell 115&     0.197&       998  \\
Abell 2163&     0.203&       1405  \\
Abell 746&     0.232&       1606   \\
%Abell 523&     0.100&       217  \\
RXC J1314.4-2515 W&     0.244&       577  \\
RXC J1314.4-2515 E&     0.244&       937  \\
Abell 521&     0.247&       692  \\
ZwCL 2341.1+0000 N*&     0.27&       1198  \\
ZwCL 2341.1+0000 S&     0.27&       767  \\
Abell 2744&     0.308&       1501  \\
MACSJ0717.5+3745*&     0.555&       298  \\
\end{tabular}
\end{center}
\label{default}
\end{table}%

\section{Simulations of radio emission from shocks}
\label{sec:simulations}

The simulations analyzed in this work were produced with the  
Adaptive Mesh Refinement code ENZO 1.5, developed by the Laboratory for Computational
 Astrophysics at the University of California in San Diego 
{\footnote {http://lca.ucsd.edu}}.
A detailed discussion on the numerical setup adopted to produce the
simulations of this work is presented in \citet{va10kp}.
Twenty galaxy clusters with masses in the range $6 \times 10^{14} \leq M/M_{\odot} \leq 3 \times 10^{15}$ were extracted from a total cosmic volume of 
$L_{\rm box} \approx 480$ Mpc/h. With the use of a nested grid approach to produce
initial conditions at high resolution in the region of cluster formation, the
final mass resolution for the Dark Matter (DM) component is $m_{\rm dm}=6.76 \cdot 10^{8} M_{\odot}$.
The refinement criterion is based on gas/DM over-density as well as 1D velocity jumps which results in a peak resolution of 25 kpc/h inside a radius of $\approx 5-6 R_{\rm v}$ ($R_{\rm v}$ begin the virial radius) from the centre of each cluster (see \citealt{va10kp} for further details).

The assumed cosmology is the ''concordance'' $\Lambda$CDM one, with $\Omega_0 = 1.0$, $\Omega_{\rm B} = 0.0441$, $\Omega_{\rm DM} =
0.2139$, $\Omega_{\Lambda} = 0.742$, Hubble parameter $h = 0.72$ and
a normalization for the primordial density power spectrum $\sigma_{8} = 0.8$. Our simulations neglect radiative cooling, star formation and AGN feedback processes. 
In this work we analyse clusters at redshifts z=0.0, z=0.3 and z=0.6 \citep[see][for complementary studies of our clusters at these epochs]{va10kp,va11turbo}.

\begin{figure}
\begin{center}
\includegraphics[width=0.49\textwidth]{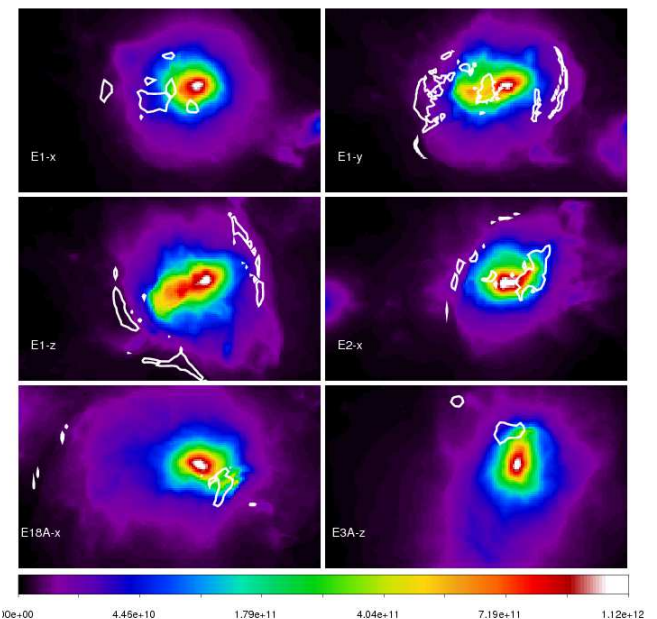}
\caption{Sample of the 6 brightest "relics" in our sample ({\it KJ\_BparallelRadial} model assumed, with $B_{\rm 0}=5\, \mu G$). The colours represent the projected X-ray bolometric emission (in arbitrary counts), the contours represent the objects identified with our relic finder. The size of each image is $3.5 \times 2.2$ Mpc/h.}
\label{fig:fig2}
\end{center}
\end{figure}

\subsection{Shock detection}

The shock waves in our runs are detected in post-processing with the algorithm presented in 
\citet{va09shocks}: this scheme is based on the analysis of 1D velocity jumps across cells. The 3D divergence of the velocity marks the centre of the shock region (see also \citealt{ry03}). 
The one-dimensional Mach number for each cell boundary is reconstructed by inverting the 1D velocity discontinuities along each scan axis using the Rankine-Hugoniot jump relations. The final Mach number is reconstructed by combining the three 1D jumps. This method has been extensively tested against similar methods used in grid codes 
\citep[e.g.][]{va11comparison}, and has proven to be
an efficient and accurate measure of shock waves in cosmological runs
\citep[see][for the statistical properties of shocks in the same sample of clusters used in this work]{va10kp}.
The co-moving kinetic power through each shock surface is given by

\begin{equation}
\label{eq:fcr}
F_{\rm KE} = \frac{\rho_{\rm u} v_{\rm s}^{3}}{2} \Delta S,
\end{equation}
where $\rho_{\rm u}$ is the co-moving up-stream density, $v_{\rm s}=M c_{\rm s}$ is the co-moving speed of the shock, $M$ is the upstream Mach number and $\Delta S$ is the surface area of the
shocked cell.

\subsection{Radio emission from shocks}

For each cluster, we generated mock radio observations along each of the three coordinate axes by creating maps of $F_{\rm KE}$, weighted
by a function, $f_{\rm M}$. This function encapsulates the poorly constrained dependence between the Mach number of the shock and the resulting radio emission. For instance, the acceleration efficiency of cosmic rays protons at shocks, the electron-to-proton energy ratio, the amplification of the magnetic field due to CR instabilities are still largely unknown \citep[e.g.][for a recent review]{br11}. Here we explore 6 different forms of $f_{\rm M}$. 

The model {\it Kin} uses $f_{\rm M}=1$, and illustrates the simple trend of
the kinetic power at shocks in the simulated volume (in this case, the acceleration efficiency of particles is independent of $M$).

In the model {\it KJ} we use the Mach number-dependent efficiency of acceleration of \citet{kj07}, $f_{\rm M}=\eta(M)$, following
the results of 1D studies of Diffusive Shock Acceleration \citep[][]{kj07} \footnote{The details of
particle acceleration in the regime of Mach numbers typical of the ICM, $2 \leq M \leq 5$
are not yet well-constrained. For instance, recent studies with particle-in-cells methods investigated
additional acceleration mechanism for particles at shocks (e.g. shock drift acceleration),
suggesting the possibilities also of a different trend with Mach number, and large efficiency for low Mach number shocks \citep{ga11}.}.

In a second set of models, we include the radial decrease of the cluster magnetic field and the 
amplification of the pre-shock magnetic field by shocks.
The radial model of reference for the "background" cluster magnetic field is the one inferred by
\citet{bo10} for observations of the Coma cluster: 
$B(\rho)=B_{\rm 0} (\rho/\rho_{\rm 2500})^{\alpha}$, with $\alpha \approx 0.5$ and $\rho_{\rm 2500}$
the density of the cluster core. 

In the model {\it KJ\_BtangledRadial}, we include a factor, $R=B_{\rm d}/B_{\rm u}= M^2/(M^2+3)$, for the amplification of an isotropic upstream magnetic field following the compression by the shock. 
%For isotropic magnetic fields, flux conservation requires that $B\propto \rho^{2/3}$ and the %synchrotron emissivity is proportional to the magnetic field energy density, $u_B\propto B^2$.

At cluster shocks, however, the amplified magnetic fields may not be isotropic and thus we also explore an alternative model.
A magnetic flux tube of strength $B_{\rm u}$, oriented at an angle $\theta_{\rm u}$ with respect to the shock normal will be bent and amplified, according to 
\begin{equation}
\label{ebk02}
B_{\rm d} = B_{\rm u} \left( \cos^2 \theta_{\rm u} + R^2 \sin^2 \theta_{\rm u} \right)^{1/2} ,
\end{equation}
where the suffices u and d refer to the up- and the downstream regions, respectively, and $R = \rho_{\rm d}/ \rho_{\rm u}$ is the shock compression ratio. From equation (\ref{ebk02}) it is easy to see that the magnetic field amplification is negligible when $B_{\rm u}$ is oriented parallel to the shock normal ($\theta_{\rm u} \sim 0$, thereby $B_{\rm d} \simeq B_{\rm u}$), whereas $B_{\rm d} \simeq R B_{\rm u}$ when $B_{\rm u}$ is nearly parallel to the shock surface.
In models {\it KJ\_BparallelRadial}, we thus assume that the amplified magnetic field at the shock is $B_{\rm d} \simeq R B_{\rm u}$.
%, such that $f_{\rm M}=\eta(M) \left (\frac{B_{\rm d}}{B_{\rm u}}\right )^2$.

In all cases the final weighting function for the kinetic energy flux is $f_{\rm M}=\eta(M) (R B(\rho))^{2}/[(R B(\rho))^{2}+B_{\rm IC}^{2}]$, where $B_{\rm IC} \approx 3.2 \cdot (1+z)^{2} \mu G$ is the magnetic field equivalent to the IC losses from CMB photons, at each redshift). In the model {\it KJ\_BtangledRadial} we
adopted the central value of $B_{\rm 0}=5\, \mu G$ as in \citet{bo10}, while in the models 
{\it KJ\_BparallelRadial} we used $B_{\rm 0}=1\, \mu G$, $=5\, \mu G$ and $=10\, \mu G$.  We assume here equal average values of the magnetic field for relaxed and very perturbed objects, which is supported by recent observations \citep[][]{bonafede11}. The formulae for $f_{\rm M}$ bracket reasonable physical models, but they cannot provide exact absolute values of the radio power in the relics.  In Sec.\ref{subsec:detection} all radio maps are normalized to the maximum within the image.

In each shocked cell of our 3D grid, we compute the energy that is thus converted to radio emission using the locally computed values of $M$, $\rho$ and $v_s$. This emission is then projected along each of the three coordinate axes, for a total of 60 mock radio images for each assumed model and at each redshift. 
 Generating projections along arbitrary angles of view \citep[e.g.][]{hoeft08} is computationally non-trivial because the computational boxes contain a large number of cells ($>500^{3}$), and, hence, we defer it to future work.
The pixels in our images have the same size ( comoving 25 kpc/h) of the maximum available 3D resolution, corresponding to the size of the primary beam of 
the Very Large Array in D configuration ($45" \times 45"$) for a luminosity distance of $\approx 220$ Mpc (z$\approx$ 0.05).

In Figure \ref{fig:fig1} we show mock radio emission maps for one simulated cluster at z=0, for 3 investigated emission
models. This cluster experienced a major merger $\sim 0.5$ Gyr ago, and two
powerful merger shocks are expanding out of the main cluster core. These shocks
are $\sim 1$ Mpc large and morphologically similar to real observed relics, and they are clearly visible in all investigated emission models. The much softer envelope of internal merger shocks and of some large-scale expanding outer ones is very difficult to detect with current radio telescopes, given that their power is several orders of magnitude dimmer than the brightest double relics, for all investigated models (see Sect.\ref{sec:results}). 

\subsection{Relic detection}
\label{subsec:detection}

The mock radio observations  at all redshifts were analysed with a 2D relic detection scheme, which identifies as part of the same relic all pixels brighter than a given threshold. Here we do not aim to reproduce the spectral and absolute power properties of relics, but only focus on their geometrical properties
in projection. We thus normalize
the pixel values of each map to the maximum within the image, and we run our 
relic finder over the $10$ per cent most luminous pixels in each map. 
In most cases, the brightest pixels are connected to form relatively large ($> 200$  comoving kpc) objects. However, depending on the threshold, some of the them can be disconnected by a few cells from the main structure. We found that by considering all pixels detached by less than 8 pixels in each direction as part of the same structure, we obtain a reasonable identification of what radio observers would typically consider a relic (tests with smaller linking lengths between pixels, 2 and 4, have yielded nearly identical results). Once we have identified all radio relics in each image and for every emission model, we determine their total power, size and distance from the centre of the host cluster (computed as the projected distance from the radio luminosity weighted centre of each relic and the X-ray peak).

In Figure \ref{fig:fig2}, we 
show the contours of the six most powerful relics detected in our sample of 60 maps at z=0
for {\it KJ\_BparallelRadial} with $B_{\rm 0}=5\, \mu G$, as detected by our algorithm. The first three panels show the emission from the same object as in Fig.\ref{fig:fig1} seen along
the three coordinate axes. In the other panels the emission comes from different simulated galaxy clusters. In all panels, we also over-plotted of the X-ray brightness map for each cluster in colours.

\begin{figure*} %
\includegraphics[width=0.32\textwidth]{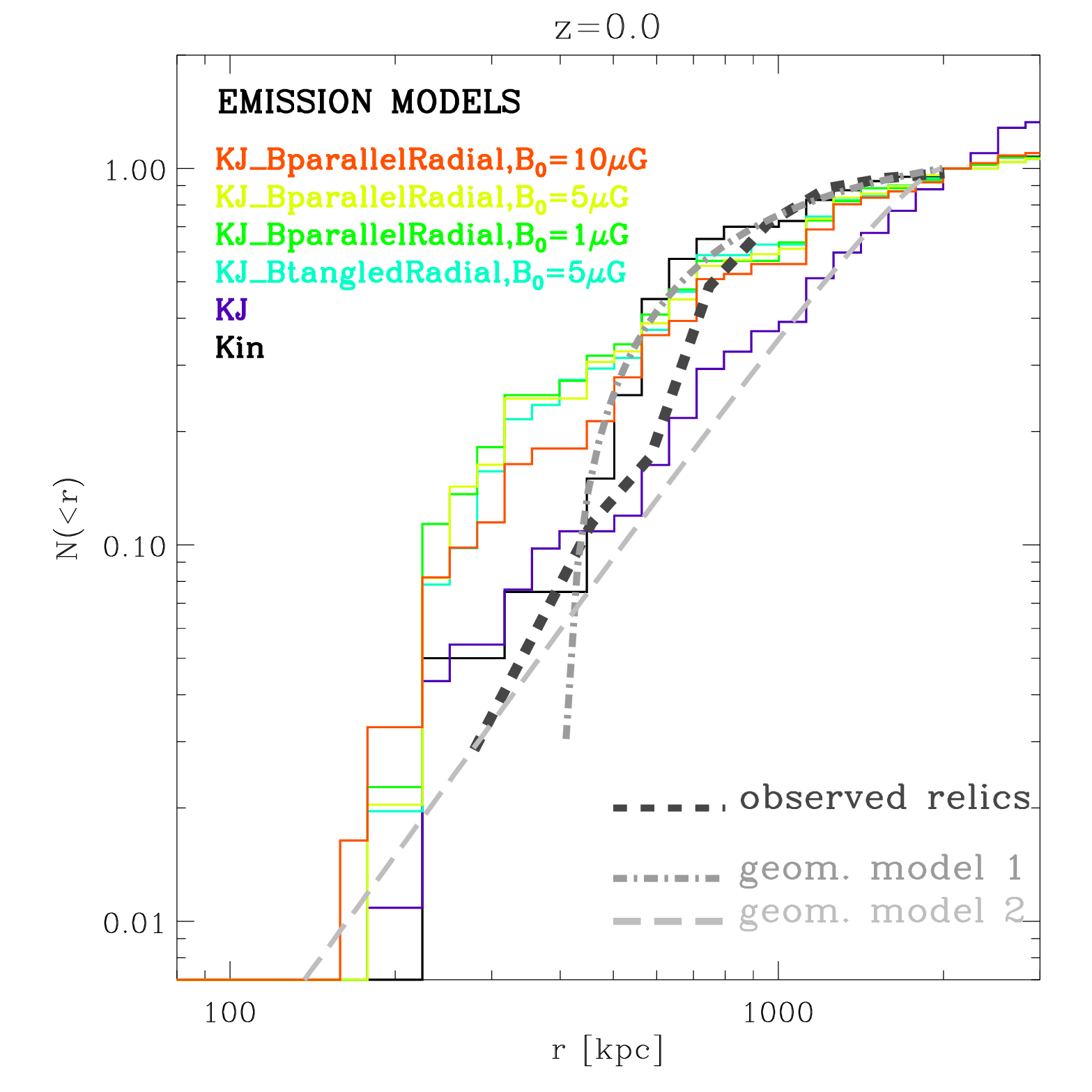}
\includegraphics[width=0.32\textwidth]{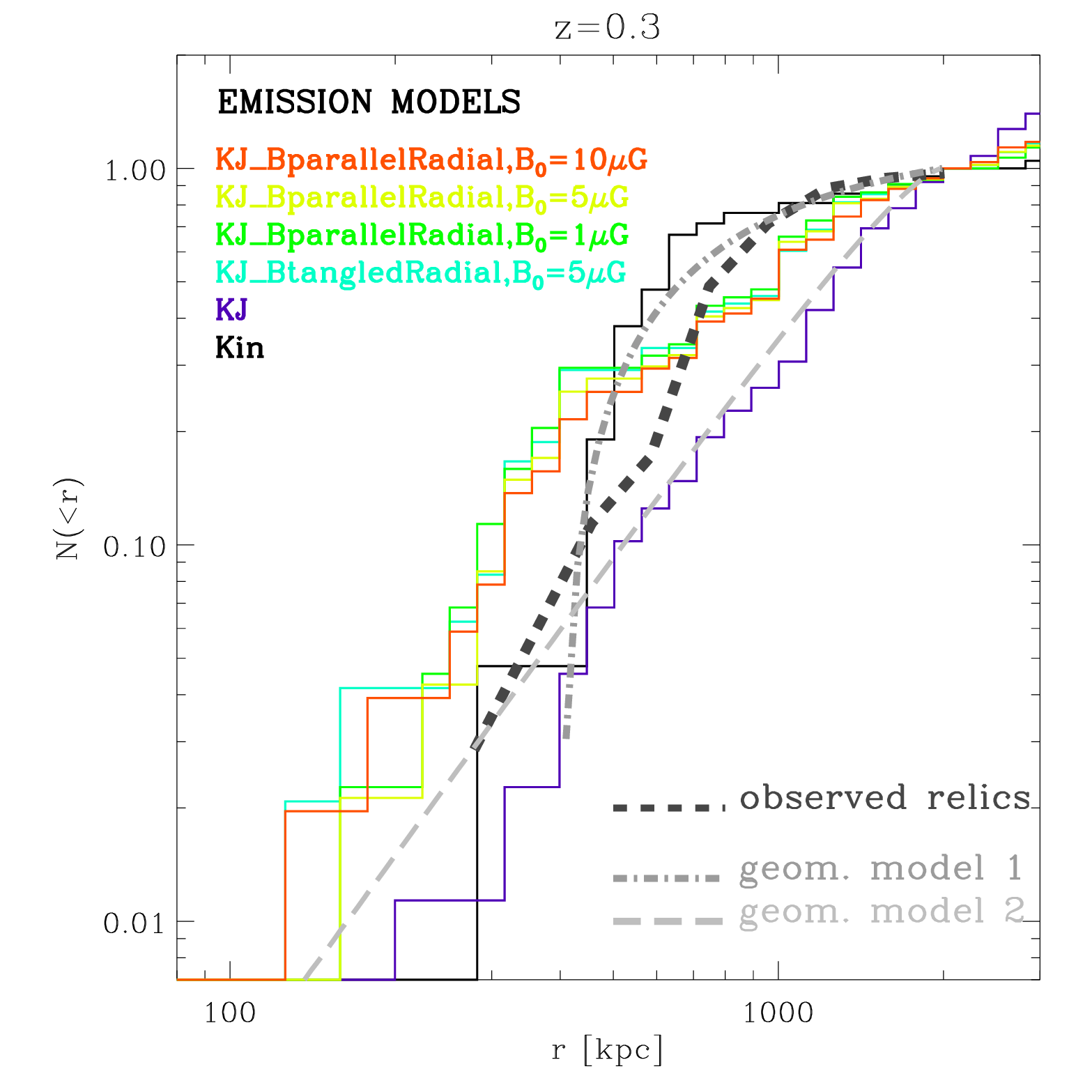}
\includegraphics[width=0.32\textwidth]{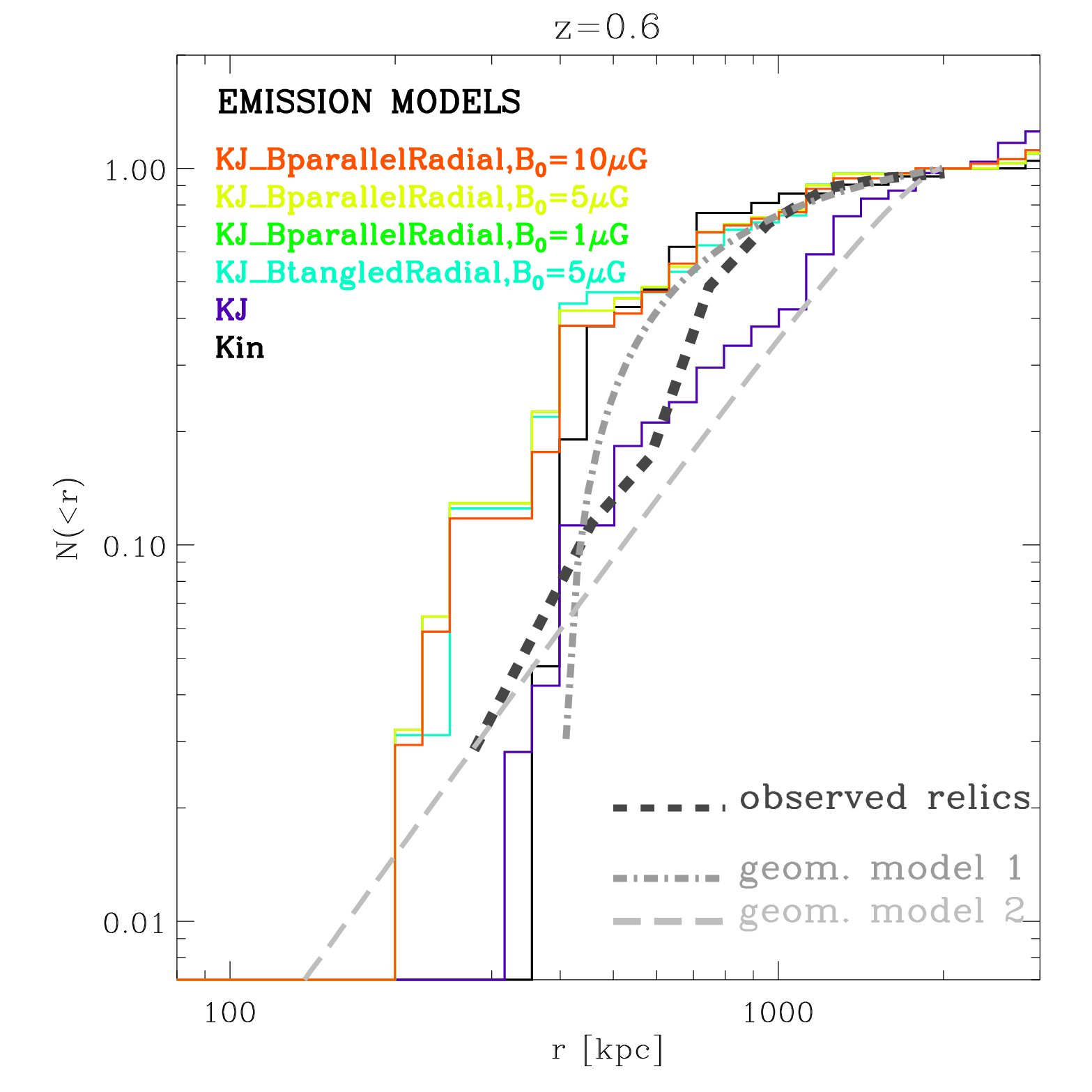}
\caption{Cumulative distributions of simulated radio gischt emissions
for the different emission models at z=0, z=0.3 and z=0.6. The
dashed dark grey line is for observed
radio relics; the light grey lines are for the geometrical models 1 and 2}, outlined in the text (see Sec.\ref{sec:results} for details).
\label{fig:fig3}
\end{figure*}

\section{Results}
\label{sec:results}

The main result of this paper is presented in Fig.\ref{fig:fig3}. Here we show the cumulative distributions of projected comoving distances from the centre for 
the relics from our sample (Table 1) as well as the radio relics detected in our mock images for 6 different emission models at the three epochs. Since we do not intend to reproduce the absolute power of relics (and their chance of  detection in mock images), we normalize all distributions to the observed one at a  comoving radius of 2 Mpc.  

In the observed distribution, $\sim 50$ per cent of relics
are located at distances $>800$ kpc from the cluster centre. 
In only one object of the sample ($\sim 3$ per cent), the 
projected radial distance is below 300 kpc (MACSJ0717.5+3745 {\footnote {As noted in Tab.1, the identification as a "radio gischt" is also debatable, and the location of the X--ray
centre of the host cluster is uncertain \citep[see][]{bonafede09, vanweeren09}.}}). At all epochs our simulated distributions show a very similar trend, with the exception of 
model {\it KJ}, in which the distributions are more extended ($\sim 50$ per cent of the relics is found at $r>1000$ kpc).  Notably, including a profile for the magnetic field distribution shifts the distribution to slightly lower radii compared to the other models. The resulting distribution matches the observed distribution in the range $r \sim 500-2000$ kpc quite well. Comparing the simulated and observed distribution at larger redshifts is more difficult, since only a few observations are available beyond z$\sim$0.3. However, Fig.\ref{fig:fig3} shows that the radial distribution
of simulated relics at higher redshifts is more radially concentrated compared to z=0, due to the fact that clusters are 
physically smaller, and that major merger are more frequent
at large redshifts. In all cases, even at higher redshifts the 
frequency of central radio relics is very low, $\leq 5$ percent
for $r \leq 300$ kpc for all investigated models.

\begin{figure} 
\includegraphics[width=0.42\textwidth]{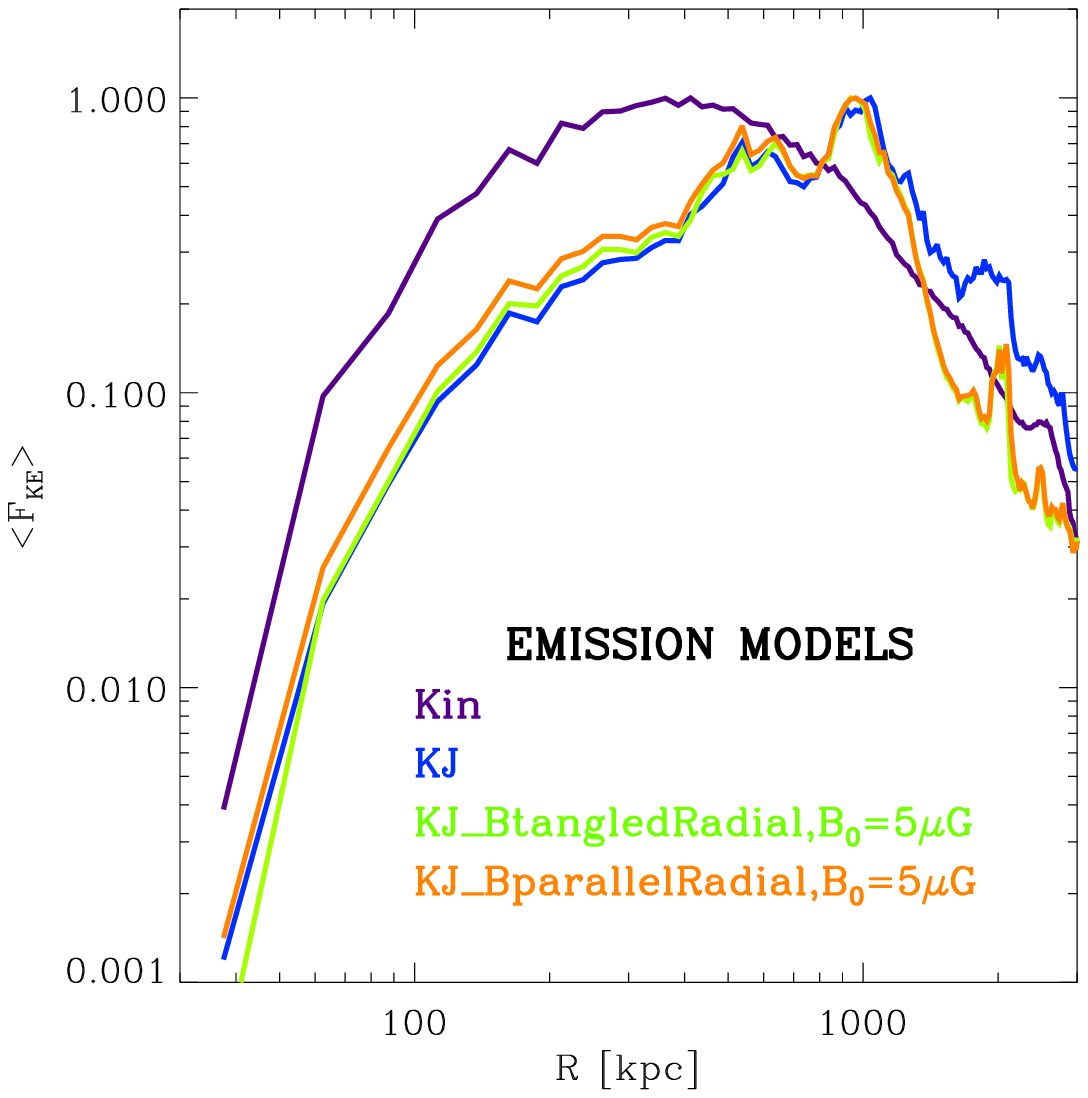}
\includegraphics[width=0.42\textwidth]{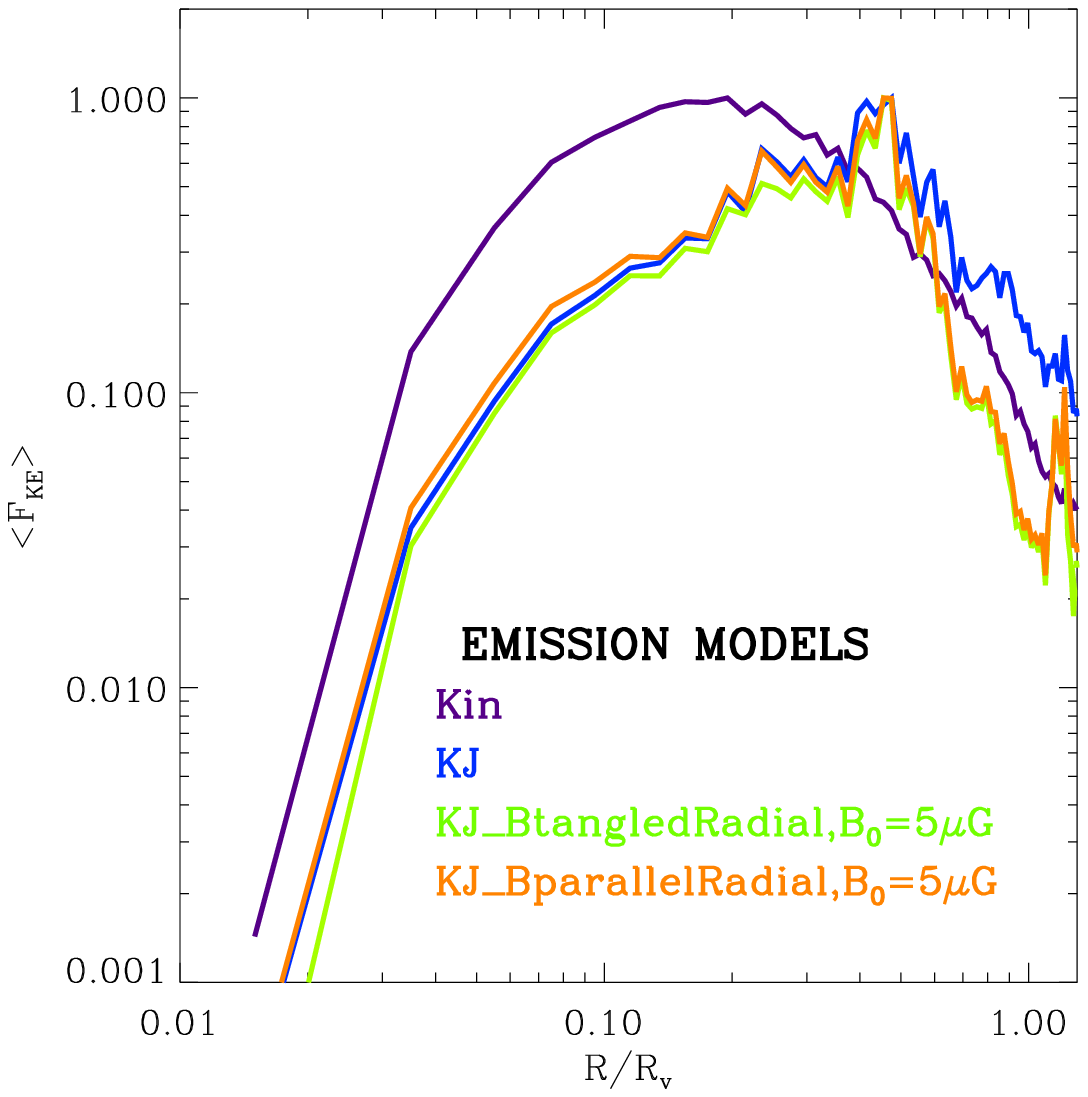}
\caption{ Volume-averaged (spherically and over all clusters at z=0) profiles of the kinetic power at shocks, $F_{\rm KE}$,
for the emission models investigated (for clarity only model {\it KJ\_BparallelRadial} with $B_{\rm 0}=5\, \mu G$ is shown). Top panel: profiles averaged at physical radii. Bottom panel: Same profiles as in the top panel, only as a function of radius normalized to the virial radius, $R_{\rm v}$, of each cluster.}
\label{fig:fig4}
\end{figure}

The observed distribution of relics for $r>800$ kpc  at z=0 is best reproduced by models that include the spatial variation of the background cluster magnetic fields, and the local amplification by shocks. Interestingly, though, the cumulative distribution is already apparent in a model where $f_{\rm M}=1$. This implies that the radial distribution of radio gischt relics is driven by the radial distribution of $F_{\rm KE}$. Hence, we have plotted radial profiles of $F_{\rm KE}$, volume-averaged  within radial shells, for all simulated clusters at z=0 (top panel of Fig.\ref{fig:fig4}). $F_{\rm KE}$ initially increases with radius, because the surface of the relics, $\Delta S$, increases roughly as $\sim r^2$, and $v_s^3$ also increases with radius because of the radial drop
of the ICM temperature outwards.  Eventually though, the decrease in $\rho_{\rm u}$ (as $\sim r^{-2}$ for a $\beta$-model outside of the cluster core, $\beta=2/3$) wins over the increase in $\Delta S$, and the profile of $F_{\rm KE}$ will turn over. The maximum of the distribution, however, depends on the assumed $f_{\rm M}$. For a fixed acceleration efficiency, the peak is 
at $\sim 400-500$ kpc, and increases to up to $\sim 1$ Mpc if the dependence of the acceleration efficiency on Mach number, $\eta(M)$, is included. In all models, we expect the number of radio relics to go down again with cluster-centric distance beyond a radius of $\sim 0.5-1$ Mpc. Since our sample has a mass range of $\sim 0.6$ dex,
we also computed the profiles normalized to the respective virial radii of the cluster (bottom panel of Fig.\ref{fig:fig4}), and found a very similar behaviour, with
maxima at $\sim 0.2-0.5 R_{\rm v}$.

To test the role played by geometrical effects, we over-plot in the same figure the result of very simple 2D geometrical model ("{\it geometrical model 1"}), which assumes that all radio relics form in a major merger with the initial size of  the core, $r_{\rm core}$ (we fix $r_{\rm core} \approx 350$ has an average value for our clusters). The radio relics increase their linear size while expanding the outer cluster atmosphere, as $l_{\rm relic}(r) \approx r-r_{\rm core}$. 
The relics are assumed to be visible only if the plane of the merger is in the plane of 
the sky, and are assumed to be invisible elsewhere due to the lack of limb brightening. In this case, the cumulative probability of finding a 
radio relic at a given distance from the main cluster centre is given by the ratio between the size of the relic at a given radius and the
circumference of the corresponding radial shell,
$N(>r)=l(r)/r=(r-r_{\rm core})/r$.  We find that this geometric model gives a reasonable description of the observed relic distribution for $r>800$ kpc (thus suggesting that the effect of limb brightening is very important for the outermost relics), but that it offers only a poor description of the observed relic distribution at smaller radii.

One may wonder why one never observes as relics the shocks in the mid-plane that form when clusters collide before core passage. These shocks turn out to have very low Mach numbers because a filament connects the two clusters. The diffuse gas in these filaments is already $>10^6$ K and this gas gets adiabatically compressed and hence heated as the two clusters approach each other. Hence, we do not find these shocks as radio relics in our cosmological simulation. 

We note that the probability of observing radio relics in projection against the 
cluster centre is very low for geometrical reasons.
Even assuming that the occurrence and properties of radio
relics are independent of distance, only a small fraction of relics
should appear projected onto the cluster centres. This follows from
simple geometrical considerations (see Suppl. Material in \citealt{br08},  "{\it geometrical model 2}" in Fig.\ref{fig:fig3}): relics projected to
within distance a $r$ from the cluster centre are
those that lie along a line-of.sight in a cylinder of radius $r$, that
intercepts a decreasing fraction of the cluster volume with radial
distance.
The resulting distribution of probability is thus:
\begin{equation}
F(<r) = {3 \over{R_{\rm 2Mpc}^3}} \int_0^{\pi/2} \sin(\theta) d\theta
\int_0^{a(\theta)} l^2 dl,
\end{equation}
where $a(\theta) = \min (r/\sin(\theta) , R_{\rm 2Mpc})$, $R_{\rm 2Mpc}=2$ Mpc.

The resulting distribution of relics is shown in Fig.\ref{fig:fig3}, the dimming of projected relics due to the lack of
limb-brightening will further decrease the chance of observing them at
small distances.

From the comparison with our simulated relics, it can be seen
that the effect of dimming further kills the probability
of detecting relics in projections on the cluster core, even
in those few cases expected by geometry. Indeed, even in the case of the most powerful relic (Fig.\ref{fig:fig1}) merely a bright spot of emission is detected at $\sim 100$ kpc from the
centre in the $y$-projection, but most of the extension of the two giant relics is lost because of the lack of limb brightening in this case. The flux goes down by the ratio between the largest linear size of relics to their widths, which is typically $>10$, \citep[e.g.][]{hoeft08,burns11}.

%In Fig.\ref{fig:fig5} we quantify this effect by plotting the isocontours for the radio emission (for the model {\it KJ\_Bparallelradial})
%along the $y$-projection of cluster E1 at z=0, showing the region where the radio emission is 0.1, 0.01 and 0.001 of the peak emission in the map. 

Based on our maps, it is likely that relics in projection against the centre of the cluster have not been detected so far because of sensitivity limits of radio observations.
When a shock is seen face-on, and the resulting relic is projected against the cluster centre,
the peak of its radio emission lies at the edge of the cluster; 
its the radio brightness would be $\sim 100$ times lower in projection onto the cluster centre. Relics have a low surface brightness ($\approx$ 1$\mu$Jy arcesc$^{-2}$ at 1.4 GHz) and apart from a few cases,  are detected at a few $\sigma$ by current radio telescopes. 
Hence, detecting central relics as those seen in our runs would require a sensitivity currently unachievable, also for future radio telescopes (such as LOFAR and SKA) because of confusion limits.

%\begin{figure} %
%\includegraphics[width=0.45\textwidth]{e1_cont.ps}
%\caption{Projected emission map of cluster E1 along the x-axis using {\it mod.5} (in log[erg/s]). The white isocontours show the region of emission
%respectively a factor 10, 100 and 1000 below the peak of emission ($\approx 6.31 \cdot 10^{35} erg/s$) in the image.}
%\label{fig:fig5}
%\end{figure}

\section{Conclusions}
\label{sec:conclusions}

In this paper we explored the question why gischt radio relics are never detected in the innermost regions of clusters.
Using simple prescriptions to model the synchrotron emission emitted at shock waves, we produced mock radio maps of a sample of 20 massive galaxy
clusters simulated at high resolution using the ENZO code,  at redshifts z=0.0, z=0.3 and z=0.6. Our models parametrized the efficiency with which electrons are accelerated at shocks, the background magnetic field and the amplification of the upstream magnetic field.
In all models, no detectable radio-emission from radio relics is found inside the innermost $\sim 100-200$ kpc of our clusters.  Nearly $\sim 50$ per cent of detectable radio relics are
located at projected distances $>800$ kpc ($>0.2-0.5 R_{\rm v}$) from the cluster centres. 
The low probability of observing radio relics close to cluster centres results from the fact that the kinetic energy dissipated in merger shock waves increases with distance from the cluster centre, combined with the more
obvious geometrical reasons. This is because the Mach number of merger shocks as well as their surface areas increase with distance from the centre faster than the density decreases. The acceleration efficiency as a function of Mach number as well as the radial dependence of the cluster magnetic field do not alter this trend qualitatively.
Finally, the lack of limb brightening makes the detection of the few central projected relics almost impossible at the typical sensitivity of all existing radio telescopes.

In summary, the lack of central radio relics in observations is due to the low energy dissipation of shock waves in cluster centres. Attempts to constrain the mechanism at work in the production of radio relics have been already explored in cosmological simulations \citep[e.g.][]{mi01,hoeft08,pf07,sk11}. In the future we also plan to study in more detail the 
statistical, spectral and morphological properties of simulated radio relics in our sample.

\section*{acknowledgements}
F.V., M.B., A.B. and K.D. acknowledge support through grant FOR1254 from the Deutsche Forschungsgemeinschaft (DFG). 
F.V. acknowledges computational resources under the CINECA-INAF 2008-2010 agreement. K.D. acknowledges the support by the DFG Cluster of Excellence "Origin and Structure of the Universe". G.B. acknowledges the support by PRIN-INAF 2009 and
ASI-INAF I/009/10/0.
We acknowledge G. Giovannini of fruitful scientific discussion, and C. Gheller for his fundamental contribution in producing the sample of simulated clusters.  Finally, we thank the anonymous referee for useful comments.

\bibliographystyle{mnras}
\bibliography{scienzo,marcus,cluster}

\begin{thebibliography}{}

\bibitem[\protect\citeauthoryear{{Bagchi} et~al.}{{Bagchi}
  et~al.}{2006}]{2006Sci...314..791B}
{Bagchi} J., {Durret} F., {Neto} G.~B.~L.,  {Paul} S., 2006, Science, 314, 791

\bibitem[\protect\citeauthoryear{{Bagchi} et~al.}{{Bagchi}
  et~al.}{2011}]{bag11}
{Bagchi} J. et~al., 2011, \apjl, 736, L8

\bibitem[\protect\citeauthoryear{{Blandford} \& {Eichler}}{{Blandford} \&
  {Eichler}}{1987}]{be87}
{Blandford} R.,  {Eichler} D., 1987, \physrep, 154, 1

\bibitem[\protect\citeauthoryear{{Bonafede} et~al.}{{Bonafede}
  et~al.}{2010}]{bo10}
{Bonafede} A., {Feretti} L., {Murgia} M., {Govoni} F., {Giovannini} G.,
  {Dallacasa} D., {Dolag} K.,  {Taylor} G.~B., 2010, \aap, 513, A30

\bibitem[\protect\citeauthoryear{{Bonafede} et~al.}{{Bonafede}
  et~al.}{2009}]{bonafede09}
{Bonafede} A., {Giovannini} G., {Feretti} L., {Govoni} F.,  {Murgia} M., 2009,
  \aap, 494, 429

\bibitem[\protect\citeauthoryear{{Bonafede} et~al.}{{Bonafede}
  et~al.}{2011}]{bonafede11}
{Bonafede} A., {Govoni} F., {Feretti} L., {Murgia} M., {Giovannini} G.,
  {Br{\"u}ggen} M., 2011, \aap, 530, A24

\bibitem[\protect\citeauthoryear{{Brown}, {Duesterhoeft}, \& {Rudnick}}{{Brown}
  et~al.}{2011}]{bdr11}
{Brown} S., {Duesterhoeft} J.,  {Rudnick} L., 2011, \apjl, 727, L25

\bibitem[\protect\citeauthoryear{{Br{\"u}ggen} et~al.}{{Br{\"u}ggen}
  et~al.}{2011}]{br11}
{Br{\"u}ggen} M., {Bykov} A., {Ryu} D.,  {R{\"o}ttgering} H., 2011, \ssr, 71

\bibitem[\protect\citeauthoryear{{Brunetti} et~al.}{{Brunetti}
  et~al.}{2008}]{br08}
{Brunetti} G. et~al., 2008, \nat, 455, 944

\bibitem[\protect\citeauthoryear{{Burns} \& {Skillman}}{{Burns} \&
  {Skillman}}{2011}]{burns11}
{Burns} J.~O.,  {Skillman} S.~W., 2011,  arXiv:1101.3361

\bibitem[\protect\citeauthoryear{{Feretti} et~al.}{{Feretti}
  et~al.}{1997}]{feretti97}
{Feretti} L., {Boehringer} H., {Giovannini} G.,  {Neumann} D., 1997, \aap, 317,
  432

\bibitem[\protect\citeauthoryear{{Finoguenov} et~al.}{{Finoguenov}
  et~al.}{2010}]{fi10}
{Finoguenov} A., {Sarazin} C.~L., {Nakazawa} K., {Wik} D.~R.,  {Clarke} T.~E.,
  2010, \apj, 715, 1143

\bibitem[\protect\citeauthoryear{{Gargat{\'e}} \& {Spitkovsky}}{{Gargat{\'e}}
  \& {Spitkovsky}}{2011}]{ga11}
{Gargat{\'e}} L.,  {Spitkovsky} A., 2011, arxiv:1107.0762

\bibitem[\protect\citeauthoryear{{Giovannini}, {Feretti}, \&
  {Stanghellini}}{{Giovannini} et~al.}{1991}]{giovannini91}
{Giovannini} G., {Feretti} L.,  {Stanghellini} C., 1991, \aap, 252, 528

\bibitem[\protect\citeauthoryear{{Hoeft} et~al.}{{Hoeft}
  et~al.}{2008}]{hoeft08}
{Hoeft} M., {Br{\"u}ggen} M., {Yepes} G., {Gottl{\"o}ber} S.,  {Schwope} A.,
  2008, \mnras, 391, 1511

\bibitem[\protect\citeauthoryear{{Kang} \& {Jones}}{{Kang} \&
  {Jones}}{2007}]{kj07}
{Kang} H.,  {Jones} T.~W., 2007, Astroparticle Physics, 28, 232

\bibitem[\protect\citeauthoryear{{Kempner} et~al.}{{Kempner}
  et~al.}{2004}]{2004rcfg.proc..335K}
{Kempner} J.~C., {Blanton} E.~L., {Clarke} T.~E., {En{\ss}lin} T.~A.,
  {Johnston-Hollitt} M.,  {Rudnick} L., 2004, in {Reiprich} T., {Kempner} J.,
  {Soker} N., ed, The Riddle of Cooling Flows in Galaxies and Clusters of
  galaxies, p. 335

\bibitem[\protect\citeauthoryear{{Miniati} et~al.}{{Miniati}
  et~al.}{2001}]{mi01}
{Miniati} F., {Jones} T.~W., {Kang} H.,  {Ryu} D., 2001, \apj, 562, 233

\bibitem[\protect\citeauthoryear{{Pfrommer} et~al.}{{Pfrommer}
  et~al.}{2007}]{pf07}
{Pfrommer} C., {En{\ss}lin} T.~A., {Springel} V., {Jubelgas} M.,  {Dolag} K.,
  2007, \mnras, 378, 385

\bibitem[\protect\citeauthoryear{{Rottgering} et~al.}{{Rottgering}
  et~al.}{1994}]{rottgering94}
{Rottgering} H., {Snellen} I., {Miley} G., {de Jong} J.~P., {Hanisch} R.~J.,
  {Perley} R., 1994, \apj, 436, 654

\bibitem[\protect\citeauthoryear{{R\"ottgering} et~al.}{{R\"ottgering}
  et~al.}{1997}]{1997MNRAS.290..577R}
{R\"ottgering} H.~J.~A., {Wieringa} M.~H., {Hunstead} R.~W.,  {Ekers} R.~D.,
  1997, \mnras, 290, 577

\bibitem[\protect\citeauthoryear{{Ryu} et~al.}{{Ryu} et~al.}{2003}]{ry03}
{Ryu} D., {Kang} H., {Hallman} E.,  {Jones} T.~W., 2003, \apj, 593, 599

\bibitem[\protect\citeauthoryear{{Skillman} et~al.}{{Skillman}
  et~al.}{2011}]{sk11}
{Skillman} S.~W., {Hallman} E.~J., {O'Shea} B.~W., {Burns} J.~O., {Smith}
  B.~D.,  {Turk} M.~J., 2011, \apj, 735, 96

\bibitem[\protect\citeauthoryear{{Skillman} et~al.}{{Skillman}
  et~al.}{2008}]{skillman08}
{Skillman} S.~W., {O'Shea} B.~W., {Hallman} E.~J., {Burns} J.~O.,  {Norman}
  M.~L., 2008, \apj, 689, 1063

\bibitem[\protect\citeauthoryear{{van Weeren} et~al.}{{van Weeren}
  et~al.}{2011}]{vw11}
{van Weeren} R.~J., {Br{\"u}ggen} M., {R{\"o}ttgering} H.~J.~A., {Hoeft} M.,
  {Nuza} S.~E.,  {Intema} H.~T., 2011, \aap, 533, A35

\bibitem[\protect\citeauthoryear{{van Weeren} et~al.}{{van Weeren}
  et~al.}{2009}]{vanweeren09}
{van Weeren} R.~J. et~al., 2009, \aap, 506, 1083

\bibitem[\protect\citeauthoryear{{van Weeren} et~al.}{{van Weeren}
  et~al.}{2010}]{vanweeren10}
{van Weeren} R.~J., {R{\"o}ttgering} H.~J.~A., {Br{\"u}ggen} M.,  {Hoeft} M.,
  2010, Science, 330, 347

\bibitem[\protect\citeauthoryear{{Vazza}, {Brunetti}, \& {Gheller}}{{Vazza}
  et~al.}{2009}]{va09shocks}
{Vazza} F., {Brunetti} G.,  {Gheller} C., 2009, \mnras, 395, 1333

\bibitem[\protect\citeauthoryear{{Vazza} et~al.}{{Vazza} et~al.}{2010}]{va10kp}
{Vazza} F., {Brunetti} G., {Gheller} C.,  {Brunino} R., 2010, \na, 15, 695

\bibitem[\protect\citeauthoryear{{Vazza} et~al.}{{Vazza}
  et~al.}{2011b}]{va11comparison}
{Vazza} F., {Dolag} K., {Ryu} D., {Brunetti} G., {Gheller} C., {Kang} H.,
  {Pfrommer} C., 2011b, arxiv:1106.2159

\bibitem[\protect\citeauthoryear{{Vazza} et~al.}{{Vazza}
  et~al.}{2011a}]{va11turbo}
{Vazza} F., {Brunetti} G., {Gheller} C., {Brunino} R.,  {Br{\"u}ggen} M.,
  2011a, \aap, 529, A17
  
\bibitem[\protect\citeauthoryear{{Venturi} et~al.}{{Venturi}
  et~al.}{2007}]{2007A&A...463..937V}
{Venturi} T., {Giacintucci} S., {Brunetti} G., {Cassano} R., {Bardelli} S.,
  {Dallacasa} D.,  {Setti} G., 2007, \aap, 463, 937

\end{thebibliography}

\end{document}